\documentclass[english,twocolumn,showpacs,preprintnumbers,amsmath,amssymb,prb]{revtex4-1}
\usepackage[T1]{fontenc}
\usepackage[latin9]{inputenc}
\usepackage{amstext}
\usepackage{graphicx}

\makeatletter

\usepackage{epsfig}

\makeatother

\usepackage{babel}
\begin{document}

\title{Critical behavior in graphene: spinodal instability at room temperature}

\author{R. Ram\'{\i}rez and C. P. Herrero}

\affiliation{Instituto de Ciencia de Materiales de Madrid (ICMM), Consejo Superior
de Investigaciones Cient\'{\i}ficas (CSIC), Campus de Cantoblanco,
28049 Madrid, Spain }
\begin{abstract}
At a critical spinodal in-plane stress $\tau_{C}$ a planar crystalline
graphene layer becomes mechanically unstable. We present a model of
the critical behavior of the membrane area near $\tau_{C}$ and show
that it is in complete agreement with path-integral simulations and
with recent experiments based on interferometric profilometry and
Raman spectroscopy. Close to the critical stress, $\tau_{C}$, the
in-plane strain behaves as $\left(\tau_{C}-\tau\right)^{1/2}$ for
$\tau<\tau_{C}$.
\end{abstract}


\maketitle
Since the first experimental characterization of graphene as a two-dimensional
(2D) one-atom thick solid membrane,\citep{novoselov04,novoselov05}
a huge amount of experimental and theoretical work has been devoted
to this material.\citep{amorim14,roldan17} The very existence of
a crystalline 2D membrane was unexpected from general symmetry arguments
by the Mermin-Wagner theorem.\citep{mermin66} The surface corrugation
of the layer was considered as an important mechanism for the modification
of its electronic properties\citep{deng15} as well as an stabilizing
factor for the planar morphology of the layer.\citep{fasolino07}

A well-known model to explain the stabilization of the planar layer
assumes that the amplitude of the out-of-plane fluctuations follows
a power-law, i.e., $\left\langle h^{2}\right\rangle \varpropto N^{1-(\eta/2)}$,
with $N$ being the number of atoms in the sheet, and $\eta$ an anomalous
exponent $\eta\sim0.8-0.85$.\citep{doussal17} The anharmonic coupling
between the out-of-plane and in-plane phonon modes increases the bending
rigidity of the layer, so that for long wavelengths the bending constant
becomes dependent on the wavevector as $\kappa(k)\propto k^{-\eta}$.
The theoretical framework for this model is the self-consistent screening
approximation (SCSA) applied to a tensionless membrane, that also
predicts that the membrane should display a negative Poisson ratio,
$\nu=-1/3$.\citep{doussal17} Several classical simulations of out-of-plane
fluctuations of graphene have been analyzed by following this theoretical
model.\citep{fasolino07,los16a,hasik18} However, to the best of our
knowledge there is no experimental confirmation that the behavior
of graphene is described by an anomalous exponent $\eta\sim0.8-0.85$.
On the contrary, there are experimental data\citep{politano12} and
computer simulations\citep{los16a} supporting that the Poisson ratio
of a graphene layer is positive $(\nu\sim0.16)$ and differs from
the predicted auxetic value of $\nu=-1/3$.

Recent analytical investigations offer an alternative explanation
for the stability of the planar morphology of the layer. By a perturbational
treatment of anharmonicity, it is predicted that free-standing graphene
displays a small but finite acoustic sound velocity in the out-of-plane
direction, caused by the bending of the layer. \citep{adamyan16,bondarev18}
Similar results were derived by different analytical perturbational
approaches.\citep{amorim14,michel15} A finite sound velocity $v$
implies that the free-standing layer displays a finite surface tension,
$\sigma=\rho v^{2},$ where $\rho$ is the density of the layer. The
surface tension $\sigma$ acts as an intrinsic tensile stress that
is responsible for the observed stability of a planar graphene layer.
Classical\citep{ramirez17} and quantum\citep{herrero18} simulations
of free-standing graphene are in excellent agreement with the theory
presented in Refs. \onlinecite{adamyan16} and \onlinecite{bondarev18}. 

Relevant physical information on the intrinsic stability of a planar
layer can be gained by studying the approach to its limit of mechanical
stability. In recent papers\citep{ramirez17,ramirez18} we have shown
that at a critical compressive in-plane stress $\tau_{C}$ a planar
graphene layer becomes mechanically unstable. At this applied stress
$\tau_{C}$, the flat membrane is unstable against long-wavelength
bending fluctuations. For $\tau>\tau_{C}$ the layer forms wrinkles,
i.e., periodic and static undulations, with amplitudes several orders
of magnitude larger than those arising from thermal fluctuations.
Such wrinkles have been often observed 
experimentally.\citep{bao09,hattab11,bao12,zhang_13,bai14,deng15,meng13}
The purpose of this work is to give a simple model of the critical
behavior of the planar layer close to $\tau_{C}$. We compare this
model with quantum simulations of a free-standing layer and confirm
its validity by the agreement to experiments that monitored the strain
of the layer through two complementary techniques: interferometric
profilometry and Raman spectroscopy.\citep{nicholl17}

Quantum path-integral molecular-dynamics (PIMD) simulations of graphene
are performed as a function of the applied in-plane stress $\tau$
at temperature $T=$ 300 K.\citep{ceperley95,herrero14} The empirical
interatomic LCBOPII model was employed for the calculation of interatomic
forces and potential energy.\citep{los05} The simulations were done
in the $N\tau T$ ensemble with full fluctuations of the simulation
cell.\citep{tu02} The simulation cell contains $N=960$ carbon atoms
and 2D periodic boundary conditions were applied. The in-plane stress
$\tau$ is the lateral force per unit length at the boundary of the
simulation cell. All results presented here correspond to a planar
(i.e. not wrinkled) morphology of the membrane. Technical details
of the quantum simulations are identical to those reported in our
previous studies of graphene and are not repeated 
here.\citep{herrero16,herrero17,herrero18,ramirez18} 

Our simulations at 300 K focus on the dependence of the membrane area
with the applied in-plane stress $\tau$. The area of the 2D simulation
cell is $NA_{p}$, $A_{p}$ being the in-plane area per atom. The
pair ($A_{p},\tau)$ are thermodynamic conjugate 
variables.\citep{fournier08,tarazona_13}
In addition, the real area $NA$ was estimated by triangulation of
the surface, which six triangles filling each hexagon of the lattice.
The six triangles share the barycenter of the hexagon as a common
vertex. The physical significance of the real area $A$ of the membrane
can be inferred from recent experiments using x-ray photoelectron
(XPS)\citep{pozzo11} and Raman spectroscopy.\citep{nicholl17} This
area $A$ is related to the average covalent CC distance while the
in-plane area $A_{p}$ yields the average in-plane lattice constant.
The difference between both has been referred earlier as the hidden
area of the membrane.\citep{nicholl15,nicholl17} An ongoing discussion
in the field of lipid bilayer membranes is that their thermodynamic
properties should be better described using the notion of a real area
$A$ rather than its in-plane projection 
$A_{p}$.\citep{fournier08,waheed09,chacon15}
The real area and the negative surface tension ($A,-\sigma$) are
a pair of conjugate variables. \citep{fournier08,tarazona_13} 

The surface tension $\sigma$ determines the long-wavelength limit
of the acoustic bending modes (ZA) of the layer. The dispersion relation
of the ZA modes in this limit can be described as\citep{ramirez16} 
\begin{equation}
\rho\omega^{2}=\sigma k^{2}+\kappa k^{4},\;\label{eq:rho_w2}
\end{equation}
where $k$ is the module of the wavevector and isotropy in the 2D
$k-$space is assumed. Numerical details of the Fourier analysis of
the amplitude of the out-of-plane atomic fluctuations to obtain the
parameters $\sigma$ and $\kappa$ from computer simulations, are
given in Ref. \onlinecite{ramirez16}. At constant temperature, the
surface tension $\sigma$ and the in-plane stress $\tau$ of the planar
layer are related as:\citep{ramirez16,ramirez17}
\begin{equation}
\sigma=\sigma_{0}-\tau,\label{eq:sigma_tau}
\end{equation}
where $\sigma_{0}$ is the surface tension for vanishing in-plane
stress. From the Fourier analysis of atomic trajectories in PIMD simulations,
one derives $\sigma_{0}\sim0.1$ N/m at 300 K.\citep{ramirez18}

For a layer made of $N$ atoms, the bending mode with largest wavelength
(or smallest $k$ module) is $k_{N}=2\pi/(NA_{p})^{1/2}$. The critical
surface tension, $\sigma_{C}$ , corresponds to the appearance of
a soft bending mode with wavenumber $\omega(k_{N})=0$. Taking into
account Eqs. (\ref{eq:rho_w2}) and (\ref{eq:sigma_tau}), 
\begin{equation}
\sigma_{C}=-\kappa k_{N}^{2}=\sigma_{0}-\tau_{C}\:.\label{eq:sigma_c}
\end{equation}
The critical surface tension displays a significant finite size effect,
$\sigma_{C}\propto N^{-1}$. It vanishes ($\sigma_{C}=0)$ in the
thermodynamic limit. Meanwhile the critical in-plane stress displays
a compressive positive value $\tau_{C}=\sigma_{0}$ in this limit.

\begin{figure}
\vspace{0.2cm}
\includegraphics[width=6.5cm]{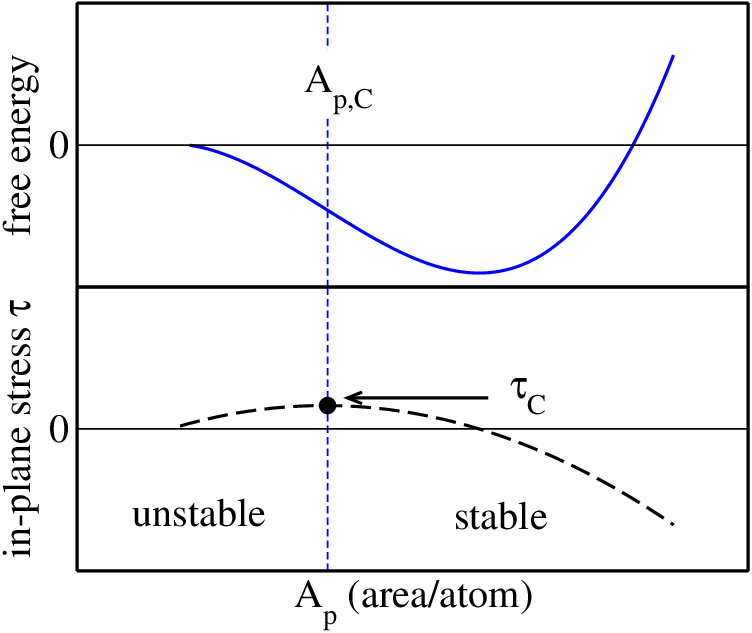}
\vspace{0.6cm}
\caption{Sketch of the dependence of the Helmholtz free energy 
and the in-plane
stress of a solid membrane as a function of its projected area. At
the critical (spinodal) point the in-plane stress takes on its maximum
compressive value, $\tau_{C}$. The spinodal point is the limit for
the mechanical stability of a planar membrane. $\tau_{C}$ displays
a size effect as described by Eq. (\ref{eq:sigma_c}) in the main
text.}
\label{fig:qualitative}
\end{figure}

The physical origin of the bending instability at $\tau_{C}$ can
be understood on a common basis with other critical phenomena in condensed
matter, e.g., cavitation of liquid helium and sublimation of noble-gas
solids under tensile stress.\citep{maris91,boronat94,bauer00,herrero03}
For any solid membrane at a given temperature the free energy $F$
depends on the in-plane surface area $A_{p}$ as displayed qualitatively
in Fig. \ref{fig:qualitative}. If a compressive stress ($\tau>0)$
is applied the in-plane area $A_{p}$ decreases. However, since the
in-plane stress $\tau=-dF/dA_{p}$ has a maximum at the inflection
point of $F$ vs $A_{p}$, there is an upper limit to the compressive
stress the planar layer can sustain. At this spinodal stress, $\tau_{C,},$
$d^{2}F/dA_{p}^{2}=0$ and from a Taylor expansion of the free energy
at the spinodal area, $A_{p,C}$, one gets for $\tau<\tau_{C}$
\begin{equation}
\tau-\tau_{C}\varpropto-(A_{p}-A_{p,C})^{2}\:.\label{eq:tau_ap2}
\end{equation}
The critical behavior of the in-plane area $A_{p}$ implies a nonlinear
stress-strain relation. Here the stress is a quadratic function of
the strain. Note that the critical in-plane stress, $\tau_{C},$ depends
on the finiteness of the graphene sample, as can be seen from Eq.
(\ref{eq:sigma_c}).

The real surface area $A$ depends on the average distance of strong
covalent CC bonds.\citep{pozzo11} The long wavelength bending of
the layer does not critically change neither the covalent distance
nor the real area $A$ of the membrane.\citep{ramirez18} One expects
here a Hooke's law: 
\begin{equation}
\tau-\tau_{C}\varpropto-(A-A_{C})\:,\label{eq:tau_a}
\end{equation}
where $A_{C}$ is the real area at the spinodal point. 

\begin{figure}
\vspace{0.1cm}
\includegraphics[width=8.0cm]{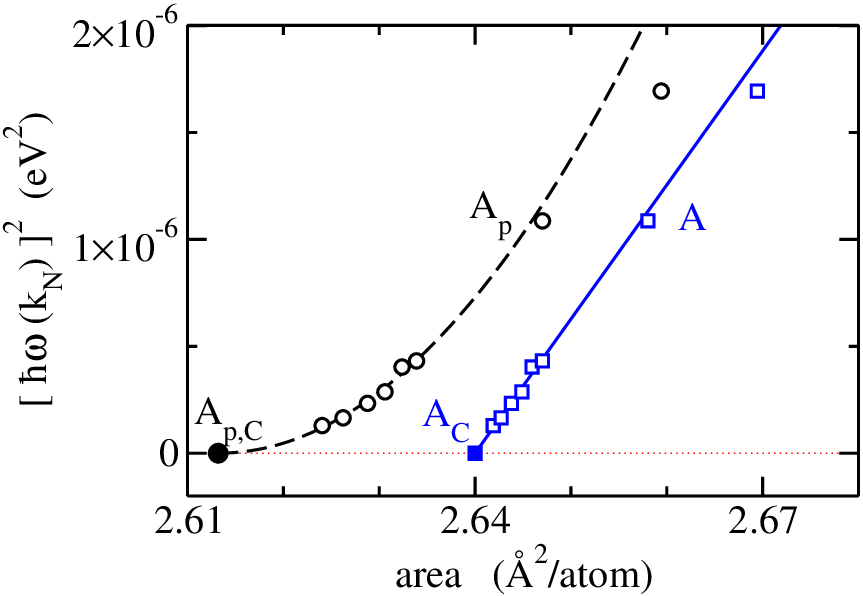}
\vspace{0.4cm}
\caption{Square of the energy quantum of the bending mode with 
wavevector $k_{N}$
vs. the in-plane (open circles) and the real area (open squares) of
the membrane at 300 K. The wavevector $k_{N}$ correspond to the bending
mode with the longest wavelength in the simulation cell. The wavenumber
vanishes $(\omega=0$) at the critical (spinodal) point. The broken
line is a quadratic fit of $\left(\hbar\omega\right)^{2}$ for $A_{p}<2.64$
$\textrm{\AA}^{2}$/atom. The full line is a linear fit of $\left(\hbar\omega\right)^{2}$
for $A<2.65$ $\textrm{\AA}^{2}$/atom. The critical areas ($A_{p,C}$
and $A_{C}$) obtained from the fits are represented as full symbols. }
\label{fig:w}
\end{figure}

The critical values $A_{p,C}$ and $A_{C}$ were obtained from the
PIMD simulations in the following way. For the wavevector with smallest
module in the simulation cell, $k_{N}$, one gets according to Eqs.
(\ref{eq:rho_w2}$-$\ref{eq:sigma_c}):
\begin{equation}
\omega{}^{2}\propto-(\tau-\tau_{C})\:.\label{eq:w_tau}
\end{equation}
The results of $\left(\hbar\omega\right)^{2}$ for the wavevector
$k_{N}$ derived at 300 K are plotted in Fig. \ref{fig:w}. The values
correspond to simulations at several in-plane stresses in the range
$0.4>\tau>-1$ N/m. The squared energies $\left(\hbar\omega\right)^{2}$
are shown as a function of the projected area $A_{p}$ (open circles),
and as a function of the real area $A$ (open squares). As the layer
is compressed, the area of the membrane and the phonon energy, $\hbar\omega$,
decrease and approach the critical point. 

At the critical (spinodal) point, the wavenumber of the bending mode
vanishes ($\omega=0$). The quadratic fit of $\left(\hbar\omega\right)^{2}$,
performed in the region where $A_{p}<$ 2.64 $\textrm{\AA}^{2}$/atom,
is displayed as a broken line in Fig. \ref{fig:w}. The vertex of
the parabola corresponds to the critical in-plane area: $A_{p,C}=2.613$
$\textrm{\AA}^{2}$/atom. The linear fit of $\left(\hbar\omega\right)^{2}$,
performed in the region $A<$ 2.65 $\textrm{\AA}^{2}$/atom, is plotted
by a full line in Fig. \ref{fig:w}. The extrapolated value of the
real area at the spinodal point is $A_{C}=2.64$ $\textrm{\AA}^{2}$/atom.
Near the critical point, $(\hbar\omega)^{2}$ varies linearly with
the in-plane stress $\tau$ {[}see Eq. (\ref{eq:w_tau}){]}. The value
of the critical stress derived from this dependence is $\tau_{C}=0.5$
N/m (see Fig. 4 of Ref. \onlinecite{ramirez18}). 

\begin{figure}
\vspace{0.3cm}
\includegraphics[width=7.0cm]{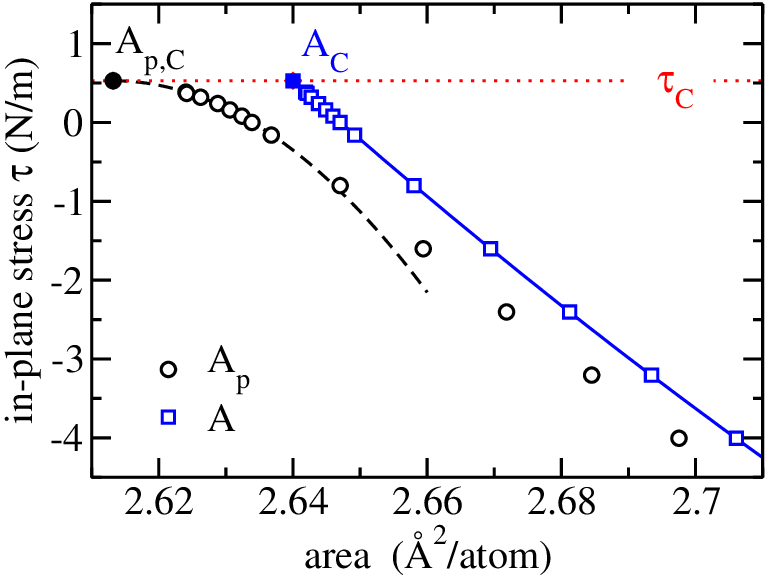}
\vspace{0.4cm}
\caption{Simulation results for the in-plane stress as a function of the
in-plane area $A_{p}$ (open circles) and as a function of the real area $A$
(open squares). Close to the spinodal tension, $\tau_{C},$ the state
equation $\tau(A_{p})$ is a parabola (broken line) with its vertex
at the critical point. For large tensile stresses ($\tau<-0.5$ N/m)
the simulation results for $\tau(A_{p})$ (open circles) lie in a
curve nearly parallel to $\tau(A)$. The full line for $\tau(A)$
is a guide to the eye.}
\label{fig:a_ap}
\end{figure}

The critical values, $\tau_{C}$ and $A_{p,C}$, are helpful data
to analyze the equation of state $\tau(A_{p})$ of graphene as derived
from the simulations. The function $\tau(A_{p})$ is displayed in
Fig. \ref{fig:a_ap} as open circles. The result resembles the sketch
displayed in Fig. \ref{fig:qualitative}. The broken line is a quadratic
f it using the critical point (closed circle) and the open circles
with $A_{p}<$ 2.64 $\textrm{\AA}^{2}$/atom. 
The critical point ($A_{p,C},\tau_{C}$)
is the vertex of the parabola. The parabola provides an excellent
description of the equation of state for stresses within the \textit{critical
region} $\tau_{C}>\tau>\tau_{C}-1\;\textrm{N/m}$. 

In Fig. \ref{fig:a_ap}, the simulation results for $\tau(A$) (open
squares) follow a linear Hooke's law, as expected from Eq. (\ref{eq:tau_a}).
At large tensile stresses ($\tau<-0.5$ N/m), i.e., outside the critical
region, the functions $\tau(A_{p})$ (open circles) and $\tau(A$)
(full line) are nearly parallel. The equation of state $\tau(A_{p})$
displays a crossover from a non-Hookean quadratic behavior in the
critical region ($\tau\gtrsim-0.5$ N/m) to a Hookean linear behavior
at larger tensile stresses ($\tau\lesssim-0.5$ N/m).

The crossover in the equation of state $\tau(A_{p})$ is a result
that should be reproduced by other simulations of graphene. In fact,
the curves $\tau(A_{p})$ derived at 300 K by classical Monte Carlo
simulations (see Fig.2 of Ref. \onlinecite{los16a}) seems to agree
with our analysis. Also recent simulations on a BN monolayer display
a critical behavior entirely similar to the one described here for
graphene.\citep{calvo_16} More important is that the equations of
state derived from the simulations, $\tau(A_{p})$ and $\tau(A)$,
can be directly compared with recent experiments. Stress-strain curves
of free-standing graphene were obtained by two complementary techniques:
interferometric profilometry and Raman spectroscopy.\citep{nicholl17}
These techniques are complementary in the sense that they are applied
to the same sample but interferometric profilometry measures the strain
$\epsilon_{Int}$ corresponding to the in-plane area $A_{p}$, while
Raman spectroscopy measures the strain $\epsilon_{Ram}$ corresponding
to the real area $A$.\citep{nicholl17} With the purpose of comparison
to the experiments, we define the linear strains from our simulation
data as
\begin{equation}
   \epsilon_{Ap}=\frac{(A_{p}-A_{p,C})}{2A_{P.C}} \: ; \: \epsilon_{A}
            = \frac{(A-A_{C})}{2A_{C}} \:.
\end{equation}
The factor 2 in the denominator converts surface into linear strain.
Here the stress is measured as the surface tension referred to its
critical value
\begin{equation}
\sigma_{rel}=\sigma-\sigma_{C}=-(\tau-\tau_{C})\:.
\end{equation}

We have considered the experimental stress-strain curves of samples
A and B of Ref. \onlinecite{nicholl17}. The graphene samples have
an unknown built-in stress. Thus the two experimental stress-strain
curves, $\epsilon_{Int}(\sigma)$ and $\epsilon_{Ram}(\sigma)$, of
a given sample have been shifted along the horizontal axis by a constant
stress. The experimental curves $\epsilon_{Int}$($\sigma)$ were
fitted to the critical relation given by Eq. (\ref{eq:tau_ap2})
\begin{equation}
\epsilon_{Int}(\sigma)=D(\sigma-\sigma_{C})^{1/2}\:,
\end{equation}
where $D$ and $\sigma_{C}$ are fitting constants. The result for
$\sigma_{C}$ is 0.16 N/m for sample A and 0.1 N/m for sample B. 

\begin{figure}
\vspace{0.2cm}
\includegraphics[width=7.0cm]{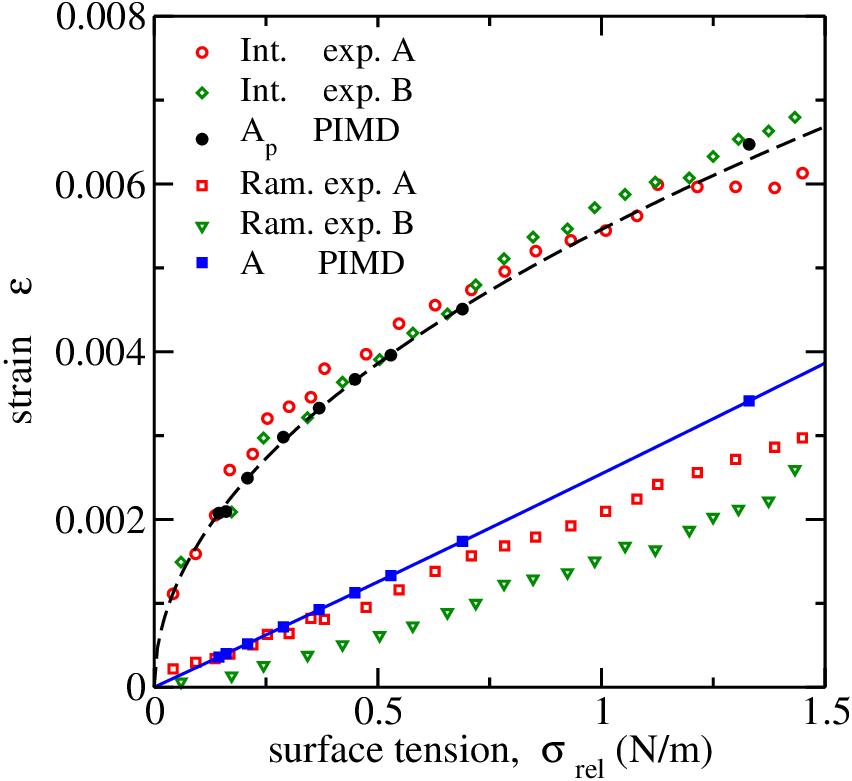}
\vspace{0.4cm}
\caption{Open symbols are experimental stress-strain curves for samples A and
B from Ref. \onlinecite{nicholl17}. The experimental strain was monitored
by interferometric profilometry (Int.) and Raman spectroscopy (Ram.)
The experimental stress of sample A was shifted by a adding a constant
of $-0.16$ N/m, and those of sample B by $-0.1$ N/m. The strains
derived from PIMD simulations are shown as closed circles for the
in-plane area $A_{p}$, and as closed squares for the real area $A$.
The broken line is the analytical stress-strain curve corresponding
to the parabola (broken line) in Fig. \ref{fig:a_ap}. The full line
is a guide to the eye.}
\label{fig:exp}
\end{figure}

The shifted experimental curves, $\epsilon_{Int}$($\sigma_{rel})$
and $\epsilon_{Ram}$($\sigma_{rel})$, for samples A and B are shown
in Fig. \ref{fig:exp} as open symbols.\citep{nicholl17} The PIMD
result for $\epsilon_{Ap}(\sigma_{rel})$ (closed circles) displays
a nearly quantitative agreement to the experimental data 
in Fig. \ref{fig:exp}.
The difference between simulation and experiment is of the same order
as the difference between the experimental results of specimens A
and B. The broken line in Fig. \ref{fig:exp} corresponds to the state
points described by the critical parabola (broken line) 
in Fig. \ref{fig:a_ap}.
The strain measured by interferometric profilometry $\epsilon_{Int}$
covers the whole critical region of the planar layer. The critical
behavior of the in-plane area is the physical explanation for the
strong nonlinearity of the experimental $\epsilon_{Int}$ curves.
In our simulations, the critical behavior of $\epsilon_{Ap}$ is solely
due to the thermal fluctuations of flexural phonons. In real graphene
devices, the presence of static wrinkles would cause an additional
increase in the measured strain $\epsilon_{Int}$. This might be a
reason to explain the stress-strain curve for a third sample C in
the experiments by Nicholl \textit{et al., }whose strain is shifted
with respect to those of samples A and B towards 
higher values.\textit{\citep{nicholl17}} 

The $\epsilon_{Ram}(\sigma_{rel})$ curves are nearly linear. The
inverse slope $d\sigma_{rel}/d\epsilon_{Ram}=2B$ is proportional
to the 2D modulus of hydrostatic compression, $B$, of the layer.
$B$ is defined by the inverse of the compressibility of the real
surface area $A$.\citep{ramirez17} The 2D compressional modulus
$B$ predicted by the employed potential model is somewhat smaller
than that derived from the experimental $\epsilon_{Ram}(\sigma_{rel})$
curves.

Summarizing, we have given a simple model of the critical behavior
of a planar graphene layer close to the compressive stress at which
it becomes unstable. The excellent agreement between stress-strain
curves derived from the model, from PIMD simulations, and from previous
experiments, provides insight into the mechanical properties of a
free-standing graphene layer. The high-quality experimental stress-strain
results of Ref. \onlinecite{nicholl17} can be quantitatively explained
by the effect of the applied stress on the equilibrium thermal fluctuations
of the layer area at room temperature. 

\acknowledgments 

This work was supported by Dirección General de Investigación, MINECO
(Spain) through Grant No. FIS2015-64222-C2-1-P. We thank the support
of J. H. Los in the implementation of the LCBOPII potential.

\bibliographystyle{apsrev4-1}
%

\end{document}